\documentclass[showpacs,aps,prb,
twocolumn
]{revtex4}
\usepackage{graphicx}
\usepackage{amsfonts}
\usepackage{amssymb}
\usepackage{amsmath}
\usepackage{color}

\begin{document}

\title{Quasi-compactons and bistability in exciton-polariton condensates}

\author{Yaroslav V. Kartashov$^{1,2}$, Vladimir V. Konotop$^3$, and Lluis Torner$^{1}$}

\affiliation{ 
$^1$ICFO-Institut de Ciencies Fotoniques, and Universitat Politecnica de Catalunya, Mediterranean Technology Park, 08860 Castelldefels (Barcelona), Spain
\\
$^2$Institute of Spectroscopy, Russian Academy of Sciences, Troitsk, Moscow Region 142190, Russia
\\
$^3$Centro de F\'isica Te\'orica e Computacional and Departamento de F\'isica, Faculdade de Ci\^encias
Universidade de Lisboa, Avenida Professor
Gama Pinto 2, Lisboa 1649-003, Portugal}
 
\date{\today}

\begin{abstract}
We address stationary patterns in exciton-polariton condensates
supported by a narrow external pump beam, and we discover that
even in the absence of trapping potentials, such condensates may
support stable localized stationary dissipative solutions (quasi-compactons), 
whose field decays faster than exponentially or even vanishes everywhere outside the pump spot. 
More general conditions lead to dissipative solitons which may display bistability.
The bistability in exciton-polariton condensates, which manifests
itself in simultaneous existence of two stable and one unstable
localized solitons with different amplitudes, widths, and
exciton-photon fractions under the same physical conditions,
strongly depends on the width of pump beam and is found to
disappear for sufficiently narrow pump beams.
\end{abstract}
%\newpage
\maketitle

%\newpage

\section{Introduction}

Triggered by a series of remarkable experiments~\cite{experiment,Balili,Amo,Amo_2}, during the last decade one observes a rapidly growing interest in study of highly populated (macroscopic) exciton-polariton states, which are also identified as quasi-particle Bose-Einstein condensates (BECs)~\cite{reviews,books}. Unlike the atomic BECs, the condensates emerging from the interaction of light and matter are nonconservative and exist due to the intense energy exchange with laser pump beam, balanced by relatively strong losses of excitons and photons arising, for example, due to material and cavity imperfections. The interplay between gain and losses makes such systems an ideal testbed for many fundamental phenomena in open quantum systems and in nonlinear science (in addition to the well appreciated practical relevance of physics of microcavities~\cite{reviews,books}).

The formation of nonlinear dissipative patterns in semiconductor microcavities may be accompanied by the bistability ~\cite{Tred,Baas04a} that arises due to Kerr-like nonlinearity induced by the exciton-exciton interactions and that was observed for extended pump beams in various setups~\cite{Baas04a,Baas04b,Shel,spinor,Sarkar,Paraiso}. The theory of this effect was developed within the framework of different meanfield models based either on equations for the polariton occupation number~\cite{Baas04b}, where the fractions of excitons and cavity photons are connected through the Hopfield coefficients~\cite{Hopfield}, or on coupled equations for excitons and photons~\cite{Ciuti}. Bistability of coherent spin ensembles described in the two-mode approximation by the spinor Gross-Pitaevskii equation with external pump has been reported too~\cite{spinor}. Under appropriate conditions bright solitons may form in semiconductor microcavities as predicted in Ref.~\cite{Yulin_bright} and observed in Ref.~\cite{Skryabin_soliton}. All such solitons, reported so far, were detected only on a broad constant background due to the use of very extended pump beams, whose width is comparable with the cavity size ~\cite{Skryabin_soliton} or considerably exceeds a width of localized excitation propagating on a broad background ~\cite{Amo}. However, in experiment the pump beam may be inhomogeneous and strongly localized in space what leads to highly nontrivial collective dynamics of polaritons within pump spot ~\cite{Wouters} and can be used for observation of rich dynamical phenomena, like direct visualization of quantum field patterns~\cite{Tosi} or switching in polaritonic transistors~\cite{Gao}.

The possibility of formation of strongly localized nonlinear excitations in condensates supported by a narrow pump beam and the impact of the width of pump beam on the localization degree of stationary states have not been explored, so far. In particular, the existence of highly localized dissipative quasi-compactons, i.e. stationary compact waves which decay faster than exponentially or in some approximation are even are exactly equal to zero everywhere except the pump spot, remains an open issue. Notice, that the concept of compacton was first introduced for continuous systems obeying nonlinear dispersion~\cite{Rosenau}. Later on compactons have been found in discrete ~\cite{disc_compact} and continuous~\cite{Abdullaev} nonlinear conservative lattices. To the best of our knowledge, dissipative compactons (representing stable attractors excitable from a broad range of initial conditions) were not detected before not only in polariton condensates, but in any dissipative system.

This work is devoted to the investigation of dissipative solitons in microcavities with localized pump. We detected the possibility of coexistence of several stable solitons with various fractions of excitons and photons in the condensate for a certain range of frequencies of pump beams, i.e. the existence of bistability, which, however, completely vanishes when the width of the pump beam becomes sufficiently small. Moreover, for the first time to our knowledge, we obtained strongly localized quasi-compactons in dissipative exciton-polariton condensates.

\section{The model}

We restrict the consideration to a quasi-one dimensional condensate of strongly coupled excitons  and photons (designated by the sub-indexes $x$ and $c$ respectively), which can be described by the Hamiltonian~\cite{Ciuti,Amo} $H=H_x+H_c+H_{nl}+H_{int}+H_p$ where $H_{x,c}$ are the Hamiltonians of the excitons and cavity photons given by
%\begin{eqnarray}
$
H_{x,c}=\int dx  \Psi_x^\dag
%h_{x,c}\left(-i\partial/\partial x\right)
\left(\hbar \omega_{x,c}- (\hbar^2/2m_{x,c})\partial^2/\partial x^2\right)
\Psi_x,
$ $ H_{nl}= (\hbar g/2)\int dx\Psi_x^\dag\Psi_x^\dag \Psi_x\Psi_x
$, describes the interaction between excitons characterized by the strength   $g>0$,
%\begin{eqnarray}
%H_c=\int dx \Psi_c^\dag \hbar \hat \omega_c\left(-i \frac{\partial}{\partial %x}\right)\Psi_c
%\end{eqnarray}
 $\,H_{int}=\hbar\Omega_R\int dx  \
\Psi_x^\dag\Psi_c+ h.c.
$ describes the interaction between light and matter which is quantified by 
the Rabi frequency $\Omega_R$,
and $H_p=\hbar\int dx
F(x)e^{i\omega_pt}\Psi_x^\dag + h.c.
$ describes pump of the photons by an external source, which is considered monochromatic and normal to the surface, i.e. $F(x)$ is considered real.

We adopt the meanfield description where the field operators $\Psi_{x,c}$ are replaced by the respective order parameters which we represent in the form $\phi_{x,c}(x,t)e^{i\omega_p t}$. Then, the system of the meanfield equations can be written in the form~\cite{Ciuti,Amo,Bloch}
 \begin{subequations}
\label{sys_dimless}
\begin{eqnarray}
\label{sys_X}
i\frac{\partial\phi_{x}}{\partial \tau}=
-
%\left(
\epsilon \phi_{x} -\frac{\kappa}{2}\frac{\partial^2\phi_{x}}{\partial \eta^2}
%\right)
%+U_{X}(x)\phi_{X}
-i\gamma_{x} \phi_{x} +\Omega\phi_{c}
%\nonumber \\
+|\phi_{x}|^2  \phi_{x}
%\end{eqnarray}
%\begin{eqnarray}
\\
\label{sys_C}
i\frac{\partial\phi_{c}}{\partial \tau}=
%\left(1
\phi_{c}-\frac{1}{2}\frac{\partial^2\phi_{c}}{\partial \eta^2}
%\right)
%+U_{C}(x)\phi_{C}
-i\gamma_{c}  \phi_{c}+\Omega\phi_{x} +f(\eta).
\end{eqnarray}
\end{subequations}
Here we introduced the dimensionless independent variables $\tau=|\Omega_c| t$ and $\eta=\sqrt{   m_c|\Omega_c|/\hbar}x$, where $\Omega_c=\omega_{p}-\omega_c$ is the pump frequency detuning from the photon frequency at $k_\|=0$ and we have limited the consideration to the negative detuning $\Omega_c<0$. We also defined the normalized pump detuning from the exciton frequency $\epsilon =(\omega_{p}-\omega_x)/|\Omega_c|$, dimensionless Rabi frequency  $\Omega=\Omega_R/|\Omega_c|$, and the relation between the masses of cavity photons and excitons $\kappa=m_c/m_x\sim 10^{-4}$. In (\ref{sys_dimless}) the normalized cavity losses of photons and the exciton decay rates are determined respectively by $\gamma_{c}$ and $\gamma_x$ and the pump shape is now desribed by $f(\eta)=F(x)/\hbar|\Omega_c|$.

The dynamics of the condensate is determined by several experimentally controllable parameters. Below as control parameters we use the amplitude $a$ and width $w$ of the Gaussian pump beam $f(\eta)=a\exp\left(-\eta^2/w^2\right)$, as well as the normalized detuning of the pump frequency $\omega_p$ from the exciton resonance, i.e. $\epsilon$ (experimentally $\epsilon$ can be varied by applying mechanical stress to the cavity that affects exciton states without changing the energy of photons~\cite{Balili}).

Due to smallness of the cavity photon mass, and thus of the coefficient $\kappa$, the kinetic energy of excitons is usually neglected~\cite{Yulin_bright}. While we do not use this approximation in numerical simulations, it is useful for qualitative discussion of the results. Indeed, by neglecting the kinetic energy of excitons in (\ref{sys_X}) one obtains the relations between the exciton and photon linear densities  $n_{x,c}=|\psi_{x,c}|^2$ and the phase difference $\theta=\theta_{c}-\theta_{x}$:
\begin{eqnarray}
\label{u_XC}
\Omega^2n_c=n_x\left[\gamma_x^2+\left(n_x-\epsilon\right)^2\right], \quad \tan\theta=\frac{\gamma_x}{n_x-\epsilon }
\end{eqnarray}
These relations imply that for $\epsilon>\sqrt{3}\gamma_x$ one can find several solutions for exciton densities for a given photon density, while for $\epsilon<\sqrt{3}\gamma_x$ for each photon density there exists a unique density of excitons.  Eqs.~(\ref{u_XC}) identify an important (real) physical parameters $\delta$ and $\varphi$, which we define by $\delta e^{i\varphi}=\epsilon-i\gamma_x$. Indeed, far from the pump spot (formally in the limit $|\eta|\to\infty$) where $n_{x,c}\to 0$ we obtain from (\ref{u_XC}) that the relation between the densities of excitons and photons is determined by $\delta$: $n_x/n_c\to\delta$, while the phase difference is fixed by $\varphi$: $\theta\to \varphi$.

\begin{figure}
\includegraphics[width=\columnwidth]{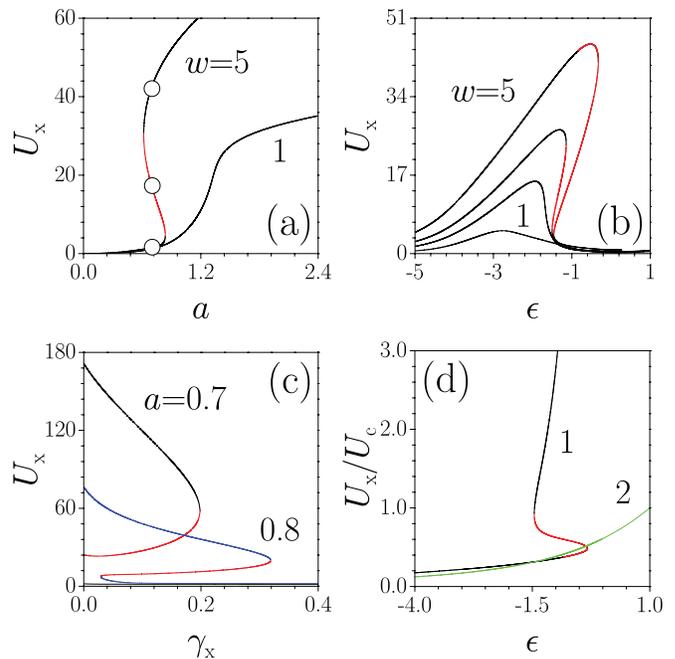}%
 \caption{Total number of excitons {\it vs}  (a) the amplitude of pump beam at $\epsilon=-1$ and $\gamma_{x,c}=0.1$ (the points correspond to solitons shown in Fig.~\ref{fig2} below), (b) the detuning $\epsilon$ at $a=0.7$  and $\gamma_{x,c}=0.1$ (the curves from top to bottom correspond to pump widths $w=5,3,2,1$), (c) the exciton losses $\gamma_x$ at $\epsilon=-1$, $w=5$, and $\gamma_{c}=0.1$. (d) The ratio $U_x/U_c$ {\it vs} frequency detuning at $a=0.7$, $w=5$, and $\gamma_{c,x}=0.1$  corresponding to soliton solution (curve 1) and the ratio $U_x/U_c$ calculated using Hopfield coefficients (curve 2). In all cases we used $\Omega=2$. The red segments correspond to unstable branches of solutions, while black segments to stable branches. 
 \label{fig1}}
\end{figure}

For inhomogeneous polariton condensates supported by a localized pump it is convenient to pass from the local densities $n_{x,c}$ to the total number of excitons and photons $U_{x,c}=\int |\phi_{x,c}|^2 dx$.  In Fig.~\ref{fig1} we show the numerically calculated number of excitons in stationary solution as a function of different parameters of the system. While for relatively wide Gaussian pump beams in certain parameter range we do observe several soliton solutions for the same value of control parameter [Fig.~\ref{fig1}(a)], the first important result of this work
 is that only one solution remains when the width of the pump beam decreases. This phenomenon is observed in the dependencies of the number of excitons on the pump beam amplitude [Fig.~\ref{fig1}(a)] and on the detuning $\epsilon$ [Fig.~\ref{fig1}(b)]. Thus, for wide pump beams the dependence $U_{x}(a)$ is "S"-shaped, while for narrow beams $U_{x}(a)$ is a monotonically growing function. Linear stability analysis of the obtained solutions predicts that only upper and lower branches of solitons in $U_{x}(a)$ dependence may be stable, while middle branch where $dU_x/da<0$   is always unstable (stable branches are shown black, while unstable branches are shown red in Fig.~\ref{fig1}). Therefore, in certain parameter range two stable solitons coexist with one unstable soliton.

While the bistability in the dependence on pump beam amplitude $a$ resembles, when it exists, the behavior of a homogeneous condensate~\cite{Ciuti,Yulin_bright}, in our setting it can also appear in the dependence on detuning $\epsilon$~\cite{comment}. Indeed, as one can observe in Fig.~\ref{fig1}(b), the exciton density is not anymore a monotonic function of this control parameter, but in the bistable region $U_x$ grows (decays) with $\epsilon$ on the upper (lower) branch. Here it is relevant to address the real physical scales at which the bistable behavior disappears. Assuming the frequency detuning $\hbar\Omega_c\sim 1\,$meV and observing that the change in the behavior occurs between $w=2$ and $w=3$ we estimate 
that bistability disappears for pump spot sizes between 3 and 5 $\mu$m. The threshold spot size can obviously be increased by increase of the detuning.

The phenomenon of bistability is also obvious in the dependence $U_{x}(\gamma_{x})$ [Fig.~\ref{fig1}(c)] that indicates that upper and middle branches merge when losses reach critical level and only one stable lower branch remains (the dependence on $\gamma_{c}$ is qualitatively similar).

In Fig.~\ref{fig1}(d) we present the ratio of the number of excitons and photons in condensate as a function of detuning $\epsilon$. Comparing with Fig.~\ref{fig1}(b) we observe that the bistability of localized polaritons is related to different exciton and photon fractions $U_x$ and $U_c$. In order to clarify the relevance of this observation we recall the fact that the exciton-polariton condensate is a mixed state of excitons and photons whose relative fractions determine the condensate properties. Affecting transmitted or reflected light the relative fraction of the quasi-particles is also relevant for interpretation of the experimental measurement~\cite{reviews}. Since the seminal work of Hopfield~\cite{Hopfield}, practically for all estimates involving these fractions the Hopfield coefficients are used. In notations of the present work, for the relative fractions of excitons $|X|^2$ and photons $|C|^2$ of the homogeneous condensate one has
 \begin{eqnarray}
\label{Hopfield}
\frac{|X|^2}{|C|^2}=\frac {\sqrt{(\epsilon-1)^2+4\Omega^2}+ \epsilon-1} {\sqrt{(\epsilon-1)^2+4\Omega^2}-\epsilon+1}.
\end{eqnarray}
This formula does not display bistability [for comparison with numerically obtained value $U_x/U_c$ in Fig.~\ref{fig1}(d) we show the relative fraction $|X|^2/|C|^2$ determined by (\ref{Hopfield})]. Since the Hopfield coefficients were used in previous studies of bistability in homogeneous exciton-polaritons BECs (see e.g.~\cite{Tred,Ciuti}), as well as in the justification of the meanfield models written for the polariton order parameter (like in the models used e.g. in~\cite{Wouters}), obviously the respective meanfield models cannot capture the effects related to variations of the relative fractions of the quasiparticles. Remarkably, Fig.~\ref{fig1}(d) shows that the branch of solutions at negative $\epsilon$ values is very well reproduced by formula (\ref{Hopfield}), while strong qualitative differences appear at larger detunings: there the exciton fraction grows significantly stronger than this is predicted by the Hopfield coefficients.

\begin{figure}
\includegraphics[width=\columnwidth]{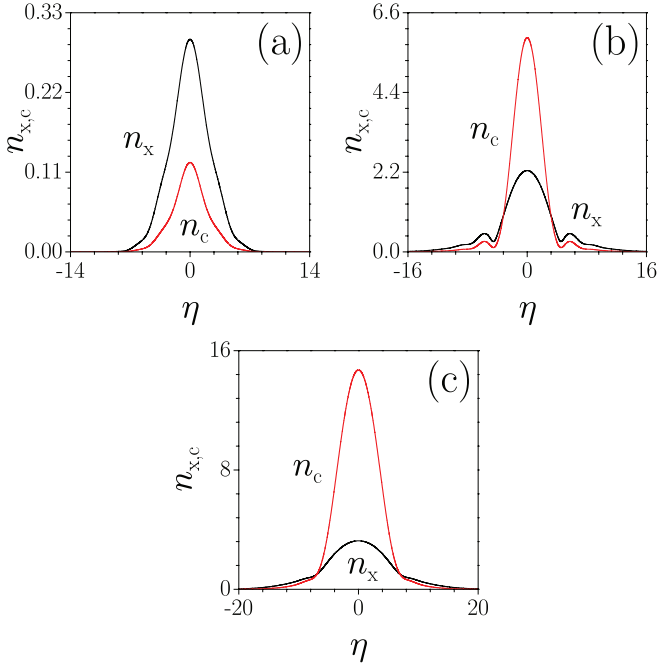}%
 \caption{The density distributions
 %for exciton $n_x$  and photon $n_c$ components
 at $a=0.7$, $w=5$, $\epsilon=-1$ and $\gamma_{x,c}=0.1$ for lower (a), middle (b), and upper (c) soliton branches corresponding to circles in Fig.~\ref{fig1}(a).
 \label{fig2}}
\end{figure}

Bistable behavior of exciton and photon fractions becomes particularly evident from typical density distributions in stationary solutions shown in Fig.~\ref{fig2} for three branches of the bistable curve from Fig.~\ref{fig1}(a) (see points marked by circles). The figure reveals several interesting features.  First, the fraction of photons is larger (smaller) than the fraction of excitons on the upper (lower) branch in Fig.~\ref{fig1}a [notice that the lower (upper) branches in panels a, b and c of Fig.~\ref{fig1} correspond to upper (lower) branch on the panel d]. Despite significant differences in fractions of photons and excitons, both components occupy nearly the same spatial domain and the widths of both components exceed the characteristic size of the pump beam.  

\begin{figure}
\includegraphics[width=\columnwidth]{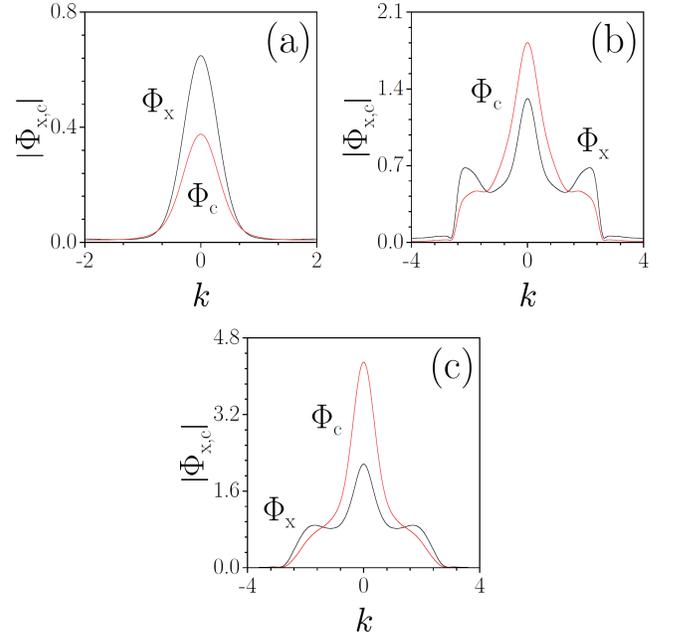}%
 \caption{Spatial spectra $\Phi_{x,c}(k)=\frac{1}{\sqrt{2\pi}}\int_{-\infty}^\infty \phi_{x,c}(x)e^{-ikx}dx$ corresponding to the profiles shown in Fig.~\ref{fig2}}
\label{fig3_0}
\end{figure}

Moreover, the soliton from lower branch (panel a) is substantially narrower than soliton from upper branch (panel c), what is reflected in the respective spatial spectra sown in Fig.~\ref{fig3_0}. Notice the appearance of oscillations on shapes and spectra of solitons from middle, i.e. unstable. branch (panel b). Existence of three spectral maxima in Fig.~\ref{fig3_0} 
indicate on the possibility of resonant mode interactions, which can be triggered by any small perturbation, what is likely to be the physical reason for the instability of the modes from the middle branch. 

\begin{figure}
\includegraphics[width=\columnwidth]{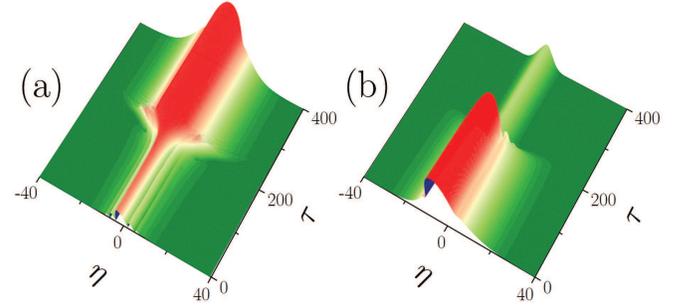}%
 \caption{Hysteresis of the dissipative solitons. (a) Switching from the lower to upper branch at $a=0.839$ (slightly exceeding the amplitude at which the lower branch ceases to exist). (b) Switching from the upper to lower branch at $a=0.613$ (that is slightly below the pump amplitude at which the upper branch ceases to exist). In both cases $w=5$, $\epsilon=-1$, $\gamma_{x,c}=0.1$, and $\Omega=2$. Only the modulus of exciton component is shown.}
\label{fig3}
\end{figure}

The existence of more than one stable state in a system with experimentally adjustable parameters implies the possibility of hysteresis. Previous theoretical
%and experimental~\cite{Baas04a,Baas04b,Paraiso}
consideration~\cite{Baas04a,Baas04b,Paraiso} dealt with hysteresis of condensates supported by nearly homogeneous pump (what corresponds to sufficiently wide pump spots in the experiment). In Fig.~\ref{fig3} we show the hysteresis of a localized dissipative soliton in the polariton condensate. In particular, Fig.~\ref{fig3}(a) shows jump from lower to upper soliton branch stimulated by small increase of the amplitude of pump beam beyond the amplitude at which lower branch ceases to exist (the input soliton corresponds to right outermost point in $U_x(a)$ dependence). Analogously, Fig.~\ref{fig3}(b) shows jump from upper to lower branch when pump beam amplitude was decreased below the threshold for existence of upper branch.

\section{Quasi-compactons}

\begin{figure}
\includegraphics[width=\columnwidth]{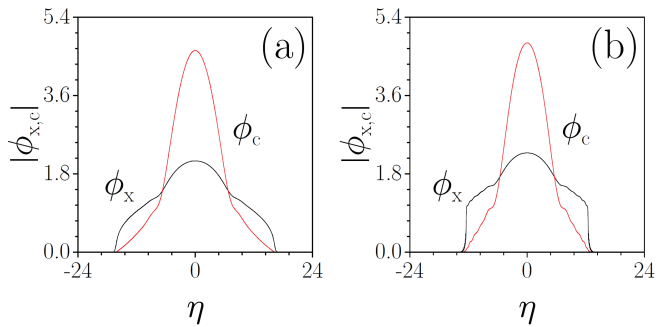}%
 \caption{The wavefunction modulus for exciton and photon components for soliton solution from the upper branch at $\epsilon=0$ (a) and $\epsilon=1$ (b) for $a=0.9$, $w=5$, $\gamma_{x,c}=0.1$, and $\Omega=2$.}
 \label{fig4}
\end{figure}

Localized soliton solutions presented here are dissipative solitons with decaying density distributions, which are supported by the localized pump beam. Somewhat similar excitations may be obtained in nonlinear optics~\cite{loc_optics}. However, the drastic difference between the case of exciton-polariton condensate considered here and optical settings is that in the case of condensate the pump is external, rather than parametric. In this case the decay law of soliton outside the pump domain may be completely different from the decay law in the free space what opens the possibilities for formation of compactons (or quasi-compactons), i.e. solutions which are zero (or displaying faster than exponential decay) outside the spot domain. In order to explain this phenomenon, we again turn to the case $\kappa=0$. From Eq.~(\ref{u_XC}) it follows that at $\eta\to\infty$ the phase and amplitude of the exciton and photon components are uniquely related by the parameters $\delta$ and $\varphi$. Hence, at $\eta\to\infty$ there exist only two free parameters defining the solution. Indeed, by neglecting the nonlinear term (which decays much faster than the linear ones) in Eq. (\ref{sys_dimless}) with $\kappa=0$ one can obtain the asymptotics of the solution at $\eta\to\infty$:
\begin{eqnarray}
\label{decay}
\phi_c\sim Ce^{(ik-\mu) \eta}-2\int_{\eta}^{\infty}\!\! f(x)\frac{\sinh((ik-\mu)|\eta-x|)}{ik-\mu}dx
\end{eqnarray}
where
\begin{eqnarray}
k^2\delta^2=\sqrt{(\delta^2+\Omega^2\epsilon )^2+(\delta^2\gamma_c+\Omega^2\gamma_x)^2}-\delta^2-\Omega^2\epsilon
\end{eqnarray}
and $\mu=(\gamma_c+\Omega^2\gamma_x/\delta^2)/|k|$ is the decay exponent. The constant $C$ is complex. Its amplitude and phase are two constants parametrizing the solution. In order to find the particular value of $C$ one has to employ the additional constrains (see e.g.~\cite{Alfimov2}). For example, for even solitons $\phi_{x,c}(\eta)=\phi_{x,c}(-\eta)$ the requirement of zero derivative of $\phi_{x,c}(\eta)$ at $\eta=0$ gives two conditions for definition of complex constant $C$ via the system parameters. If $f(\eta)$ is nonzero and decays faster than exponentially (as in the present work) two situations are possible: if system parameters are such that $C\neq 0$, the decay of solution is exponential, but if $C=0$ the solution is still nonzero, but it decays in accordance with the law determined by the shape of the pump beam $f(\eta)$, i.e. faster than exponentially. Such solutions are dissipative quasi-compactons and their shapes obtained from Eqs.~(\ref{sys_dimless}) are presented in Fig.~\ref{fig4} for the case of $\kappa=10^{-4}$. 
Quasi-compactons appear only on the upper and middle soliton branches in the region of positive detuning, when $\epsilon>0$ (i.e. when $\omega_p>\omega_x$). Typically, the larger the value of  $\epsilon$ the faster is the decay of tails of solution. From Fig.~\ref{fig1}(b) one can see that upper soliton branches, hence quasi-compacton solutions, can be found in the region with positive detuning only for sufficiently large pump width $w$.  Quasi-compactons can be stable.

Remarkably, not only quasi-compacton, but also exact dissipative compacton solutions can be obtained by choosing proper pump beam shapes in the limit $\kappa=0$. To illustrate this, let us withdraw the requirement for the pump $f(\eta)$ to be real and choose
\begin{eqnarray}
\label{f}
f(\eta)=
-f_0\delta e^{-i\varphi}+f_0[(3-i\gamma_c)\delta e^{-i\varphi}+\Omega^2]\cos^2\eta
\nonumber \\
+\Omega^2f_0^3\cos^4\eta\left[15-(19-i\gamma_c)\cos^2\eta\right],
\end{eqnarray}
for $|\eta|<\pi/2$ and $f=0$ otherwise  ($f_0$ is a free parameter, also recall that $\delta e^{i\varphi}=\epsilon-i\gamma_x$). Then, the exact compacton solution of Eq.~(\ref{sys_dimless}) with $\kappa=0$ reads
\begin{equation}
\label{compacton}
\phi_x =-f_0 \Omega  \cos^2\eta
    ,\,\,\,
    \phi_c =f_0\cos^2\eta \left(f_0^2\Omega^2\cos^4\eta-\delta e^{-i\varphi}  \right)
\end{equation}
for $|\eta|<\pi/2$ and $\phi_x\equiv \phi_c\equiv 0$ otherwise. We confirmed numerically that this compacton solution is structurally (i.e. at $\kappa=10^{-4}$) and dynamically stable. We also notice that the suggested solution is not unique one of such kinds. As a matter of fact a large diversity of the dissipative modes with compact support (i.e. equal exactly zero beyond a give region) can be constructed, although by using even more sophisticated gain profiles. 

Explicit analytical solution of a dissipative compaction allows to address one more  important issue, which is the dissipative nature of the obtained solutions. To this end let us first set $\gamma_c=\gamma_x=0$. Then (\ref{f}), (\ref{compacton}) still hold valid and represent a Hamiltonian system subjected to an external force $f(\eta)$. Such a system can be viewed as a composed quasi-particle with one component (excitonic) having an infinite mass (since $\kappa=0$) affected by the inhomogeneous force applied to the photonic component. This force is balanced by the force of interaction between the components (determined by $\Omega$) giving origin to the stationary state described by (\ref{compacton}) with  $\gamma_c=\gamma_x=0$.
It turns out, however, that stationary localized stationary solutions of Eq.~(\ref{sys_dimless}) can be found only when at least one of the parameters $\gamma_{x,c}$ is nonzero. In this case one can obtain one or three soliton solutions depending on the particular value of pump amplitude, as shown in Fig.~\ref{fig1}(a). When both $\gamma_x=\gamma_c=0$ stationary states are unstable and one observes considerable radiation from pump region.

\section{Conclusion}

Summarizing, we showed that spatially localized pump supports strongly localized dissipative solitons in exciton-polariton condensates. The width of the pump beam determines whether the system possesses or is free from bistability. We found that in the bistability regimes the relative fractions of excitons and photons may significantly differ from the predictions for the homogeneous condensate where they are determined by the Hopfield coefficients.  

Although bistability is considered to be a common feature for externally driven systems (it occurs not only in exciton-polariton condensates, but also in lasers, as well as in various systems described by driven nonlinear Schrödinger equation [30]), in real experiments it is usually manifested in the abrupt jump of the norm (or power) of the output state with increase of the pump amplitude, so that observation of localized states from certain power range usually cannot be performed due to their instability. In this respect, the obtained here fact that by simple change of the size of the pump spot the states from the above mentioned power range can be stabilized is rather interesting.

For the first time to our knowledge we presented dissipative compactons, or more precisely quasi-compactions, i. e. solutions which decay faster than exponentially. 
%We notice that recently an interest on modes supported by localized gain of a finite dimension was %raised in the context of optical applications~\cite{loc_optics} (see also~\cite{book_chapter} for %a brief review of the available results). 
Thus it is of interest further study of quasi-compactons from the point of view of driving such modes by laser beams, of their interactions, their use for switches and observation of the  hysteresis phenomenon, as well as many other problems typical for the nonlinear physics.

Finally, it should be stressed that the consideration in this work was based on the   meanfield model consisting of the coupled equations for the excitons and photons, thus leaving a series of open questions. Among them we mention  the extension of the theory for spinor model of the condensate;  analysis of possible excitations of the condensation in excited states and associated energy relaxation~\cite{Wouters2}; extension to two-dimensional geometry; dynamical properties and possibility of managing localized excitations, etc.

\acknowledgments

VVK is grateful to A. V. Yulin and to B. Pietka for valuable discussions and comments. The work of VVK was supported by the FCT (Portugal) grants PTDC/FIS/112624/2009 and PEst-OE/FIS/UI0618/2011.


\begin{thebibliography}{99}
\bibitem{comment} While $\epsilon$ is not the most typical control parameter used in the experiments, as we already mentioned its variation can be achieved by applying mechanical stress~\cite{Balili}.  On the other hand by changing other control physical parameters, like for example the detuning $\Omega_c$, will lead to simultaneous change of the dimensionless variables. This, however, will affect the reported effects only quantitatively not changing them qualitatively.
\end{thebibliography}
\end{document}